\documentclass[11pt]{article}
\usepackage{amssymb}
\usepackage{bm}
\newcommand{\ket}[1]{\left | #1 \right \rangle}
\newcommand{\bra}[1]{\left \langle #1 \right |}

\def\openone{\leavevmode\hbox{\small1\kern-3.8pt\normalsize1}}
\def\tr{{\rm tr}\; }
\def\ce{{\cal E}}

\def\cb{{\cal B}}
\def\ch{{\cal H}}
\def\cn{{\cal N}}

\def\cm{{\cal M}}

\def\cp{{\cal P}}
\def\ct{{\cal T}}
\def\cv{{\cal V}}

\def\acc{{\rm Acc}}

\def\CC{\mathbb{C}}

\newtheorem{theorem}{Theorem}

\newtheorem{lemma}{Lemma}

\newtheorem{proposition}{Proposition}
\newtheorem{corollary}{Corollary}

\newcommand{\proj}[1]{\ket{#1}\!\bra{#1}}

\newcommand{\inner}[2]{ \langle #1 | #2 \rangle}

\newcommand{\beq}{\begin{equation}}
\newcommand{\eeq}{\end{equation}}
\newcommand{\beqa}{\begin{eqnarray}}
\newcommand{\eeqa}{\end{eqnarray}}

\newcommand{\qed}{\mbox{\rule[0pt]{1.5ex}{1.5ex}}}

\begin{document}
\begin{center}
{\LARGE\bf Entanglement cost of generalised measurements }\\
\bigskip
{\normalsize  Richard Jozsa$^\dagger$, Masato Koashi$^{\ddag
\dagger}$, Noah Linden$^\S$,  Sandu Popescu$^\natural$,\\ Stuart
Presnell$^\dagger$,
Dan Shepherd$^\sharp$ and Andreas Winter$^\dagger$}\\
\bigskip
{\small\it $^\dagger$Department of Computer Science,
University of Bristol,\\ Merchant Venturers Building, Bristol BS8 1UB U.K.
\\[1mm]

$^\S$Department of Mathematics, University of Bristol,\\
University Walk, Bristol BS8 1TW U.K.\\[1mm]

$^\ddag$School of Advanced Studies, The Graduate University for
Advanced Studies \\ (SOKENDAI),  Shonan Village, Hayama, Kanagawa, 240-0193, Japan\\[1mm]

$^\natural$Department of Physics, University of Bristol,
\\ Tyndall Avenue, Bristol BS8 1TL, U.K.\\[1mm]

$^\sharp$CESG, PO Box 144, Cheltenham, Gloucestershire, GL52 5UE
U.K.}
\\[4mm]
\date{today}
\end{center}

\begin{abstract} Bipartite entanglement is one of the fundamental
quantifiable resources of quantum information theory. We propose a
new application of this resource to the theory of quantum
measurements. According to Naimark's theorem any rank 1
generalised measurement (POVM) $M$ may be represented as a von
Neumann measurement in an extended (tensor product) space of the
system plus ancilla. By considering a suitable average of the
entanglements of these measurement directions and minimising over
all Naimark extensions, we define a notion of entanglement cost
$E_{\min } (M)$ of $M$.

We give a constructive means of characterising all Naimark
extensions of a given POVM. We identify various classes of POVMs
with zero and non-zero cost and explicitly characterise all POVMs
in 2 dimensions having zero cost. We prove a constant upper bound
on the entanglement cost of any POVM in any dimension. Hence the
asymptotic entanglement cost (i.e. the large $n$ limit of the cost
of $n$ applications of $M$, divided by $n$) is zero for all POVMs.

The trine measurement is defined by three rank 1 elements, with
directions symmetrically placed around a great circle on the Bloch
sphere. We give an analytic expression for its entanglement cost.
Defining a normalised cost of any $d$-dimensional POVM by $E_{\min
} (M)/\log_2 d$, we show (using a combination of analytic and
numerical techniques) that the trine measurement is more costly
than any other POVM with $d>2$, or with $d=2$ and ancilla
dimension 2. This strongly suggests that the trine measurement is
the most costly of all POVMs.

\end{abstract}


\section{Introduction}\label{intro}
Bipartite entanglement is one of the fundamental quantifiable
resources of quantum information theory. In this paper we will
propose a new application of this resource to the theory of
quantum measurements, motivated by the following considerations.

Let $\cm = \{ A_\mu:A_\mu\geq 0, \sum A_\mu =I \}_{\mu=1}^m$ be a
positive operator valued measure (POVM) or ``generalised
measurement'' on the space $\ch_d$ of dimension $d$. We will be
interested in POVMs with rank 1 elements $A_\mu = \proj{k_\mu}$,
where the $\ket{k_\mu}$'s are sub-normalised and generally
non-orthogonal vectors. In quantum theory only orthogonal states
have the fundamental property of being reliably distinguishable
and in a measurement the outcomes correspond to classical
information which is similarly reliably distinguishable. This
connection between orthogonality and measurement outcomes is
embodied in the definition of a (complete) von Neumann
measurement, as being based on mutually orthogonal directions. For
a (rank 1) generalised measurement the vectors $\ket{k_\mu}$ are
not orthogonal yet the outcome information is again classical. In
this case the appearance of orthogonality is restored by Naimark's
theorem \cite{peres,naimarkthm} which asserts that any generalised
measurement may be viewed as a von Neumann measurement on a
suitable enlarged space comprising the target system together with
an auxiliary system or ancilla. In the present work we wish to
think of von Neumann measurements and their in-built orthogonality
as the fundamental substrate of all quantum measurement. The
formalism of (non-orthogonal) POVMs is then merely a mathematical
artifice that expresses the residual effect on the system when we
perform an orthogonal measurement in an extended setting. We will
see in section \ref{formulation} below that the adjoining of an
ancilla to represent $\cm$ as a von Neumann measurement provides a
natural way of associating an entanglement cost to the measurement
$\cm$. For direct von Neumann measurements on the target system
(where no ancilla is required), the entanglement cost will be
zero.

A possible motivation or application of such a notion of
entanglement cost for a generalised measurement arises in the
study of information acquisition from a quantum source. Let $\ce
=\{ \ket{\psi_i}; p_i \}$ be a source of pure quantum signal
states $\ket{\psi_i}$ with prior probabilities $p_i$. Write
$\Pi_i= \proj{\psi_i}$. If $\cm = \{ A_\mu:A_\mu\geq 0, \sum A_\mu
=I \}_{\mu=1}^m$ is any POVM on the signal space then its
measurement on a signal state results in a joint probability
distribution $p(i \& \mu)= p(\mu|i)p(i)= (\tr A_\mu \Pi_i) p_i$.
We can then calculate the associated mutual information $I(\cm:\ce
)$ which is interpreted as the amount of information that the
measurement outcome provides about the identity of the state
$\ket{\psi_i}$. For a given source $\ce$ we are often interested
in determining $\cm$ so that $I(\cm :\ce)$ is maximised.  It is
known \cite{davies} that for any source an optimal measurement may
be regarded as having all rank 1 elements. Indeed in generality
suppose that $A_\mu$ has spectral decomposition $A_\mu = \sum
\lambda_i \proj{\lambda_i}$ and in the measurement $\cm$ we
replace each $A_\mu$ by the set of rank 1 operators $\{ \lambda_i
\proj{\lambda_i} \} $, giving a refined measurement $\cm'$ that is
rank 1. Then for any source $\ce$ we have $I(\cm' :\ce) \geq I(\cm
:\ce )$. Thus it is natural to restrict attention to only rank 1
POVMs, with elements of the form $A_\mu = \proj{k_\mu}$ and we
have
\[ p(\mu |i)= |\inner{\psi_i}{k_\mu}|^2 \hspace{1cm} p(\mu)=\sum_i
p_i |\inner{\psi_i}{k_\mu}|^2. \]

It is known that some sources have a direct von Neumann
measurement as an optimal measurement but there exist other
sources for which no such direct von Neumann measurement is
optimal. For example the binary source $\cb = \{
\ket{\psi_1},\ket{\psi_2};p_1,p_2 \}$ always has an optimal von
Neumann measurement \cite{levitin} whereas the so-called trine
source of three symmetric equiprobable states in two dimensions,
defined by  $\ct =\{ \ket{\alpha_1},
\ket{\alpha_2},\ket{\alpha_3}; 1/3,1/3,1/3 \}$ with \[
\ket{\alpha_j}=\cos \frac{j\pi}{3}\ket{0}+\sin
\frac{j\pi}{3}\ket{1}, \] has the property that no von Neumann
measurement is optimal \cite{holevo}. Let $\ket{\beta_i}$ be three
vectors orthogonal to $\ket{\alpha_i}$ respectively and
subnormalised to have lengths $\sqrt{2/3}$. Then it may be shown
\cite{sasakietc} that the POVM $\cm_\ct = \{ \proj{\beta_i}
\}^3_{i=1} $ is an optimal measurement for $\ct$.

Thus we see that some sources (such as $\ct$ having no direct von
Neumann measurement as optimal) require an investment of
entanglement to reveal the full quota of their available
information, while others (such as any binary source) require no
such entanglement input. Intuitively (for $\ct$) the information
needs to be ``spread out'' in a larger space in order for it to
become ``fully visible''. This framework of ideas leads to many
interesting new questions. For example we can ask: which source
(of all qubit sources, say) requires the most entanglement
investment to see its information content? We will study these
issues at a slightly more fundamental level, focussing on the
generalised measurements themselves, rather than the sources,
using the precise formulations given in the next section.

This paper is organised as follows. In section \ref{formulation}
we develop a precise formulation of our notion of entanglement
cost for generalised measurements. Our adopted approach is based
on the formalism of Naimark extensions which we briefly review. In
section \ref{naimark} we give a constructive characterisation of
all possible Naimark extensions of a given POVM which provides a
basis for developing further properties of our notion of
entanglement cost in section \ref{properties}: firstly we identify
various classes of POVMs with zero and non-zero cost, and
explicitly characterise all POVMs in 2 dimensions having zero
cost. Secondly we derive an analytic expression for the
entanglement cost of the trine measurement (defined above).
Thirdly we derive various upper bounds on the entanglement cost
including a constant upper bound (independent of the dimension)
for any POVM.  Next, in section \ref{mostcostly} we consider the
question of which POVM has the greatest cost. Introducing the
normalised entanglement cost of $\cm$ (in $d$ dimensions) as the
cost of $\cm$ divided by $\log_2 d$ (i.e. the cost per qubit of
source space) we show that the trine measurement is more costly
than any other POVM with $d\geq 3$ and conjecture that it is the
most costly of all POVMs. In section \ref{asymptcost} we introduce
the notion of the {\em asymptotic} entanglement cost of a POVM
(defined as the large $n$ limit of the cost of $n$ applications of
$\cm$ divided by $n$) and show that it is zero for all POVMs.
Finally in section \ref{etc} we make some further remarks on the
significance of our definition of entanglement cost and briefly
discuss various other possible approaches.

   \section{Formulation of the problem}\label{formulation}

    In our definition of entanglement cost, Naimark's theorem will
    play a fundamental role so we begin by elaborating its statement here. Let
    $\ch_d$ denote the signal space (of dimension $d$) and $\ch_e$ the
    ancilla space (of dimension $e$). Let $\ket{a_0}$ be any fixed
    chosen state of the ancilla. Let $\cm = \{ A_\mu \}_{\mu=1}^m$ be
    any rank 1 POVM and write $p(\mu |\psi )=\bra{\psi}A_\mu
    \ket{\psi}$ for the probability of obtaining outcome $\mu$ on the
    state $\ket{\psi}\in \ch_d$. Naimark's theorem states that $\cm$
    may always be realised as a von Neumann measurement in $\ch_d
    \otimes \ch_e$ for $e$ sufficiently large to have $de\geq m$, in
    the following sense. Let $\ket{\psi,a_0}=\ket{\psi}\ket{a_0}$ be
    the state obtained from $\ket{\psi}$
    by adjoining the ancilla in its standard state.
    Then there is an orthonormal basis $\{ \ket{w_\nu}: \nu=1, \ldots
    ,ed \}$ in $\ch_d\otimes \ch_e$ such that for all $\ket{\psi}$:
    \[ p(\nu |\psi)=|\inner{\psi,a_0}{w_\nu}|^2 \hspace{5mm}
    \mbox{when $\nu=1, \ldots ,m$}. \] Any such orthonormal basis
    (for any choice of $\ch_e$) is called a Naimark extension of
    $\cm$. Any generalised measurement admits many inequivalent
    Naimark extensions and in section \ref{naimark} below we will see
    how to classify and construct them all. Since probabilities must
    sum to 1 it follows that \begin{equation}\label{zero} p(\nu
    |\psi)=|\inner{\psi,a_0}{w_\nu}|^2 =0 \hspace{5mm} \mbox{for
    $m<\nu \leq ed$} \end{equation} i.e. the POVM is represented in
    the first $m$ dimensions of the von Neumann measurement $\{
    \ket{w_\nu} \}$ and then the remaining von Neumann measurement
    directions must be orthogonal to the subspace $\ch_d \otimes
    \ket{a_0} \subseteq \ch_d\otimes \ch_e$. These latter
    measurement directions lie in the span of $\{ \ch_d \otimes
    \ket{a_i}: i\geq 1 \}$ where $\{ \ket{a_0}, \ket{a_1}, \ldots
    \ket{a_{e-1}} \}$ is an orthonormal extension of $\ket{a_0}$ in
    $\ch_e$. (More generally if the POVM is not rank 1 then Naimark's
    theorem states that it may similarly be represented as a von
    Neumann measurement that is incomplete i.e. not rank 1).

    The states $\ket{w_\nu}$ are generally entangled across the
    decomposition $\ch_d\otimes \ch_e$ and their entanglements will
    form the basis of our definition of entanglement cost for $\cm$.
    Note that if $\cm$ is a von Neumann measurement in $\ch_d$ then
    $m=d$, so we can take $e=1$ and set
    $\ket{w_\nu}=\ket{k_\nu}\ket{a_0}$. Thus our entanglement cost
    will be zero but for more general POVMs we will obtain non-zero
    values.

    In some treatments of Naimark's theorem (such as \cite{peres}) the
    basic signal space $\ch_d$ is enlarged not via a tensor product
    $\ch_d\otimes \ch_e$ but instead via an embedding into a larger
    dimensional space $\ch_d \hookrightarrow \ch_N$ of suitable size
    to accommodate the number of elements of the POVM. For us it is
    important that the extension arises from a tensor product and
    indeed, from a physical point of view the adjoining of an
    additional ancilla system is the only available physical means of
    extending a space while retaining the original system intact.

Turning now to the formulation of our problem,
   we begin with the following ingredients: (i) a source $\ce = \{
   \ket{\psi_i};p_i \}$ in $\ch_d$; (ii) a rank 1 POVM $\cm = \{
   \proj{k_\mu} \}_{\mu=1}^m$ with $m$ outcomes, on $\ch_d$; (iii)
   any Naimark extension $\cn = \{ \ket{w_\nu} \}_{\nu=1}^{de} $ of
   $\cm$, which is an orthonormal basis of $\ch_d\otimes \ch_e$.
Let $E_\nu$ denote the entanglement of $\ket{w_\nu}$ (defined as
the von Neumann entropy of the reduced state of $\ch_d$ or
$\ch_e$) and let
\begin{equation}\label{pnu} p(\nu)=\sum_i p_i p(\nu |i) = \sum_i
p_i |\inner{\psi_i a_0}{w_\nu}|^2 = \sum_i p_i
|\inner{\psi_i}{k_\nu}|^2
\end{equation} be the posterior probability of obtaining outcome
$\nu$ on $\ce$ (where in the last expression we take
$\ket{k_\nu}=0$ for $\nu >m$). Then introduce $E=\sum_\nu p(\nu)
E_\nu$, the average entanglement of the Naimark extension
directions, weighted by their posterior probability of occurrence
on $\ce$. We define the entanglement cost of $\cm$ on $\ce$ to be
$E_{\rm min}$, the minimum of $E$ over all choices of Naimark
extension of $\cm$.

Note that if $m$ (the number of POVM elements) is not an integer
multiple of $d$ then any Naimark extension will involve extra
directions not associated with any POVM element. However eq.
(\ref{zero}) shows that the entanglements of these extra states
$\ket{w_\nu}: \nu >m$ do not enter into $E$. A further desirable
feature of our definition is the fact that any von Neumann
measurement will have entanglement cost zero (as noted
previously).

Our definition of $E_{\rm min}$ depends on both $\cm$ and $\ce$.
To obtain a characteristic of a generalised measurement alone we
note that the POVM condition $\sum \proj{k_\mu} =I$ implies $\sum
\inner{k_\mu}{k_\mu}=d$ so that $\{ q_\mu =\inner{k_\mu}{k_\mu}/d
\}_{\mu=1}^m$ is a probability distribution canonically defined by
$\cm$. From eq. (\ref{pnu}) we see that $p(\nu)=\sum_i p_i
\inner{k_\nu}{\psi_i}\inner{\psi_i}{k_\nu}$ will equal $q_\nu$ if
$\sum_i p_i \proj{\psi_i}=\frac{1}{d}I$ so we can interpret the
canonical distribution $q_\mu$ as the posterior probability of
obtaining outcome $\mu$ on any {\em maximally mixed} source i.e.
any source having $\sum_i p_i \proj{\psi_i}=\frac{1}{d}I$ (e.g.
the trine source for $d=2$).

Using these ingredients we can state our central problem:\\ {\bf
Problem:} Given a POVM $\cm = \{ \proj{k_\mu} \}_{\mu=1}^m$ on
$\ch_d$ let $q_\mu= \frac{1}{d}\inner{k_\mu}{k_\mu}$. Let $\{
\ket{w_\nu} \}$ on $\ch_d\otimes \ch_e$ be any Naimark extension
of $\cm$ and let $E_\nu$ be the entanglement of $\ket{w_\nu}$. Let
$E=\sum_{\mu=1}^{m} q_\mu E_\mu$. Then we wish to determine
$E_{\rm min}(\cm)$, the minimum value of $E$ over all choices of
Naimark extension of $\cm$. $\qed$

Note that our definition of entanglement cost (based on the
entanglements of the Naimark extension states) is actually a
measure of entanglement {\em production} by the POVM, rather than
a measure of ``input'' entanglement cost, necessary to realise the
POVM in some suitable sense. Further discussion of this point will
be given in section \ref{others} below.

\section{Characterising all Naimark extensions of $\cm$}
\label{naimark}

In \cite{peres} Naimark's theorem is proved by showing how to
construct a Naimark extension of any rank 1 POVM. However the
extension of the space is achieved by subspace embedding $\ch_d
\hookrightarrow \ch_N$, rather than via a tensor product,
adjoining an ancilla. Below we adapt this method to work in the
system--plus--ancilla formalism (c.f. also \cite{preskill}) and
show further that it provides a constructive classification of all
possible Naimark extensions of the POVM.

Let $\cm =\{ \proj{k_\mu} \}_{\mu=1}^m$ be any rank 1, $m$-element
POVM in $\ch_d$. Let the orthonormal basis $\{ \ket{w_\nu}
\}_{\nu=1}^{de}$ in $\ch_d\otimes \ch_e$ be any Naimark extension
of it. The action of the POVM on states $\ket{\psi}$ in $\ch_d$
corresponds to the Naimark extension acting on
$\ket{\psi}\ket{a_0}$ which in turn requires a choice of ancilla
state $\ket{a_0}$. When we need explicit mention of the chosen
ancilla state we will refer to a ``Naimark extension with basic
ancilla state $\ket{a_0}$''.

We begin by identifying a relationship between the vectors
$\ket{k_\mu}$ and $\ket{w_\mu}$. If $\{ \ket{a_i} \}_{i=0}^{e-1}$
is any orthonormal basis of $\ch_e$ that extends $\ket{a_0}$, we
can write
\[ \ket{w_\nu} = \ket{v_\nu^0}\ket{a_0}+ \ket{v_\nu^1}\ket{a_1} +
\ldots + \ket{v_\nu^{(e-1)}} \ket{a_{(e-1)}} \] for suitable
(sub-normalised) coefficient vectors $\ket{v_\nu^i}$ in $\ch_d$.
The Naimark extension condition is $|\inner{\psi , a_0}{w_\nu}|^2=
|\inner{\psi}{k_\nu}|^2$ for $\nu=1,\ldots ,m$ and $|\inner{\psi ,
a_0}{w_\nu}|^2=0$ for $\nu >m$. Thus we have \[
|\inner{\psi}{v^0_\nu}|= |\inner{\psi}{k_\nu}| \hspace{5mm}
\mbox{for all $\ket{\psi}$ and $\nu=1, \ldots ,m$} \] and by lemma
\ref{lem1} below we get
$\ket{v_\nu^0}=e^{i\theta_\nu}\ket{k_\nu}$. These phases do not
affect the entanglement that we are studying and without loss of
generality we can take $\theta_\nu=0$ (e.g. they can be absorbed
into overall phase choices for the $\ket{w_\nu}$'s). Indeed, the
phase of $\ket{k_\nu}$ is not even defined by the POVM.
\begin{lemma} \label{lem1} If $|\inner{a}{x}|=|\inner{b}{x}|$ for all $\ket{x}$
then $\ket{a}=\ket{b}e^{i\theta}$.\end{lemma}

\noindent {\bf Proof}\,\, Putting $x=a$ and $b$ gives
$|\inner{a}{b}| = \inner{a}{a}=\inner{b}{b}$ and then the
projection of $\ket{a}$ perpendicular to $\ket{b}$ is zero. $\qed$

\noindent For $\nu>m$ we have $\inner{\psi , a_0}{w_\nu}=0$ so
$\ket{v_\nu^0}=0$ for $\nu >m$. Hence we have proved:

\begin{proposition}\label{num1} If $\{ \ket{w_\nu}
\}_{\nu=1}^{de}$ is any Naimark extension of $\{ \proj{k_\mu}
\}_{\mu=1}^m$ with basic ancilla state $\ket{a_0}$ then the
$\ket{w_\nu}$ have the form
\[ \begin{array}{rclll} \ket{w_\nu} & = & \ket{k_\nu}\ket{a_0}+ &
\ket{v^1_\nu}\ket{a_1}+\ldots +\ket{v_\nu^{(e-1)}}\ket{a_{(e-1)}}
&
\mbox{for $\nu=1, \ldots ,m$} \\
\ket{w_\nu} & = &  & \ket{v^1_\nu}\ket{a_1}+\ldots
+\ket{v_\nu^{(e-1)}}\ket{a_{(e-1)}} & \mbox{for $\nu=m+1, \ldots
,de$} \end{array} \] i.e. $\ket{v^0_\nu}=\ket{k_\nu}$ for $\nu =1,
\ldots ,m$ and $\ket{v^0_\nu}=0$ for $\nu>m$.
\end{proposition}

\noindent We also have a converse result:

\begin{proposition}\label{num2} Let $\{ \ket{w_\nu}
\}_{\nu=1}^{de}$ be any orthonormal basis of $\ch_d\otimes \ch_e$.
Using the basis $\{ \ket{a_i} \}_{i=0}^{e-1}$ write \[
\ket{w_\nu}= \sum_{i=0}^{e-1} \ket{v^i_{\nu}}\ket{a_i}. \] Then
$\{ A_\nu=\proj{v^0_\nu} \}_{\nu=1}^{de}$ is a POVM on $\ch_d$.
Similarly for each $i$, $\{ A_\nu^{(i)}=\proj{v^i_\nu}
\}_{\nu=1}^{de}$ is a POVM on $\ch_d$.\end{proposition}

\noindent {\bf Proof}\,\, For any $\ket{\xi}$ in $\ch_d$ we have
$|\inner{\xi , a_0}{w_\nu}|^2=|\inner{\xi}{v_\nu^0}|^2$ so
\[ \sum_\nu \inner{\xi}{v_\nu^0}\inner{v_\nu^0}{\xi}=1. \]
Writing $P=\sum_\nu \proj{v_\nu^0}$ we thus have
$\bra{\xi}P\ket{\xi}=1$ and $P$ is also Hermitian. Taking
$\ket{\xi}$ as the eigenstates of $P$ we see that all eigenvalues
must be 1 so $P=I$ and $\{ \proj{v^0_\nu} \}_{\nu=1}^{d}$ is a
POVM. Similarly by first applying a unitary swap operation on
$\ch_e$ that interchanges the roles of $\ket{a_0}$ and $\ket{a_i}$
we get the corresponding result for each $i$. $\qed$

Returning to our question of constructing all possible Naimark
extensions of $\cm =\{ \proj{k_\mu} \}_{\mu=1}^m$, proposition
\ref{num1} shows that we need to characterise all possible sets
$\ket{v_\nu^i}$ for $i\geq1$ which make the $\ket{w_\nu}$'s
orthonormal. This is achieved as follows. We will work in terms of
components with respect to a chosen basis $\{ \ket{e_j} \}$ of
$\ch_d$ and $\{ \ket{a_i} \}$ of $\ch_e$. Let $M$ be the $m \times
d$ matrix whose $m$ {\em rows} are the components of the
$\ket{k_\mu}$'s for $\mu=1, \ldots ,m$.

\begin{proposition}\label{num3} The POVM condition $\sum_\mu A_\mu =I$
is equivalent to the following condition: the $d$ {\em columns} of
$M$ are an orthonormal set of vectors in $\CC^m$.\end{proposition}

\noindent {\bf Proof}\,\, We have $\ket{k_\mu}=\sum_j M_{\mu
j}\ket{e_j}$ so the POVM condition $\sum_\mu \proj{k_\mu}=I$ is
\[ \sum_\mu M_{\mu j}M_{\mu k}^* = \delta_{jk} \] i.e. for each $j$
and $k$ the $j^{\rm th}$ and $k^{\rm th}$ columns of $M$ are
orthonormal. $\qed$

We will write any vector $\ket{\Phi}\in \ch_d\otimes \ch_e$ as a
row vector of components lexicographically ordered in the product
basis $\ket{e_j}\ket{a_k}$. Thus we have $de$ components and if \[
\ket{\Phi}=\sum_{i=0}^{e-1} \ket{v^i}\ket{a_i} \] then the first
$d$ numbers in the row vector are the components of $\ket{v^0}$,
the next $d$ are the components of $\ket{v^1}$ and so on; the last
$d$ are the components of $\ket{v^{(e-1)}}$.

Above we introduced the $m\times d$ matrix $M$ (having the
$\ket{k_\mu}$'s as rows). Let $Z$ be the $(de-m)\times d$ matrix
with all zero entries. Then by proposition \ref{num3} the columns
of \[ P= \left[  \begin{array}{c} M\nonumber \\ Z \nonumber
\end{array} \right] \] are $d$ orthonormal vectors in $\CC^{de}$.
We can extend these $d$ columns to a full orthonormal basis  of
$de$ vectors in $\CC^{de}$ by adjoining further columns to get a
$de\times de$ matrix
\[ L= \left[  \begin{array}{cc} M & A \nonumber \\ Z & B \nonumber
\end{array} \right]. \]
Here $A$ has size $m\times (de-d)$ and $B$ has size $(de-m)\times
(de-d)$. The freedom in the choice of this extension is the set of
all maximal orthonormal families in $\CC^{de}$ that are orthogonal
to the first $d$ columns of $L$. This set is isomorphic to the set
of all orthonormal bases of $\CC^{de-d}$.

\begin{proposition}\label{num4} For any square matrix $R$,
$R$ has orthonormal columns iff $R$ has orthonormal
rows.\end{proposition}

\noindent {\bf Proof}\,\, This is a standard fact from linear
algebra. $\qed$

Now $L$ has orthonormal columns so by proposition \ref{num4} $L$
has orthonormal rows too. (Indeed $L$ is a unitary matrix.) Hence
the rows of $A$ and $B$ read out in segments of length $d$ give
the vectors $\ket{v_\nu^i}$, $i\geq1$, of a Naimark extension of
$\cm$ (as in propositions \ref{num1} and \ref{num2}). Conversely
we see that every Naimark extension arises in this way: taking any
Naimark extension as in proposition \ref{num1} we write out the
$de\times de$ matrix $L'$ with the components of the
$\ket{w_\nu}$'s as rows. Then the rows of $L'$ are orthonormal so
by proposition \ref{num4} the columns are orthonormal too. Also by
proposition \ref{num1} the entries of $L'$ in the positions of $Z$
are all zero.

Hence we have a prescription for generating all possible Naimark
extensions of the POVM $\{ \proj{k_\mu} \}_{\mu=1}^m$ in dimension
$d$ with an ancilla of dimension $e$:
\begin{itemize}
\item Write down the matrix $M$ whose rows are the components of
the vectors $\ket{k_\mu}$.
\item Extend the columns of $M$ by zeroes to get a total length of $de$.
\item Extend these $d$ orthonormal columns to a full orthonormal basis of $\CC^{de}$
(by adjoining further columns) giving a square matrix $L$.
\item Look at the rows of $L$ to read off the Naimark extension states.
 \end{itemize}

 \subsection{Bounding the ancilla size}\label{ancbound}

 Although unboundedly large ancillas (i.e. $e$ values) can be used
in the above prescription, they do not help in reducing the
entanglement cost of $\cm$. It suffices to consider only $e\leq
md$ as shown next. \begin{theorem}\label{thm1} For the $m$ element
POVM
 $\cm = \{ \proj{k_\mu} \}_{\mu=1}^m$ let $\{ \ket{w_\nu} \}_{\nu=1}^{dn}$
be any Naimark extension with ancilla dimension $n$ and
entanglement cost $E$. Then there exists a Naimark extension of
$\cm$ with ancilla dimension $e\leq md$ which also has
entanglement cost $E$.\end{theorem}

\noindent {\bf Proof}\,\, Each $\ket{w_\nu}$ is an entangled state
in $\ch_d\otimes \ch_n$ and hence has support of dimension at most
$d$ in the ancilla space. So in the ancilla space at most $md$
dimensions are utilised by the $m$ states $\ket{w_\nu}$ for
$\nu=1,\ldots ,m$. Hence by a local unitary transformation on the
ancilla (which does not affect the entanglements), we can rotate
these dimensions into the first $md$ positions. Hence an ancilla
of dimension $md$ suffices to allow a Naimark extension with the
same entanglement cost. $\qed$

\section{Some properties of the entanglement cost of $\cm$}
\label{properties}
\subsection{Measurements with zero and nonzero entanglement cost}
In this subsection we characterise a class of POVMs for any
dimension $d$, with zero entanglement cost and also a class which
has non-zero cost. For $d=2$ we give a full characterisation of
zero cost POVMs.

\begin{proposition}\label{subadd} Let $\cm_0 = \{ \proj{k_\mu}
\}_{\mu=1}^m$ and $\cm_1 = \{ \proj{l_\nu} \}_{\nu=1}^n$ be two
POVMs. If $\{ p_0, p_1 \}$ is a probability distribution define
the convex combination POVM (with $m+n$ elements) $\cm = \bigcup_i
p_i \cm_i = \{ p_0 \proj{k_\mu}, p_1 \proj{l_\nu}
\}_{\mu=1,\nu=1}^{m,n}$. Then \[ E_{\min } (\cm) \leq \sum_i p_i
E_{\min } (\cm_i). \] \end{proposition} {\bf Proof}\,\, Let $B_0 =
\{ \ket{w^0_\mu} \}$ and $B_1=\{ \ket{w^1_\nu} \}$ be Naimark
extensions of $\cm_0$ and $\cm_1$ respectively, that realise the
minimal entanglement costs. We can enlarge the smaller ancilla
space so without loss of generality we may assume that the ancilla
spaces have the same dimension $e$ and that the fundamental state
$\ket{a_0}$ (c.f. proposition \ref{num1}) coincides for $B_0$ and
$B_1$. Now consider the set $B=\{ \ket{w^0_\mu}\ket{0},
\ket{w^1_\nu}\ket{1} \}_{\mu=1,\nu=1}^{me,ne}$ where $\ket{0}$ and
$\ket{1}$ are orthonormal states of a new system which we view as
an extension of the ancilla. This is clearly an orthonormal set
and if we take the fundamental ancilla state $\ket{a_0'}=\ket{a_0}
(\sqrt{p_0} \ket{0}+\sqrt{p_1}\ket{1})$ then $B$ is a Naimark
extension of $\cm$. Since the ancilla in each state of $B$ has
been extended as a product state, the entanglements are unchanged
so the entanglement cost of $\cm$ is at most $\sum_i p_i E_{\min }
(\cm_i)$. $\qed$

\noindent Note that in the above construction, if $\cm_0$ and
$\cm_1$ have elements which are proportional, they remain as
separate elements in the convex combination POVM. Since von
Neumann measurements have zero cost we immediately get:
\begin{corollary} \label{mixvonn}
Let $\cv_1, \ldots , \cv_K$ be von Neumann measurements and $\cp =
\{ p_1, \ldots ,p_K\}$ a probability distribution. Then the POVM
$\cm = \bigcup_i p_i \cv_i$ has zero entanglement cost.
\end{corollary}

$\cm$ in the corollary has the physical interpretation of a
probabilistic mixture of von Neumann measurements i.e. we first
take a classical sample from $\cp$ and then perform the
corresponding von Neumann measurement. If $\cm = \{ \proj{k_\mu}
\}_{\mu=1}^m$ is any POVM on qubits (i.e. $d=2$) then we may
represent the elements $\proj{k_\mu}$ as subnormalised Bloch
vectors $v_\mu$ of length $r_\mu=\sqrt{\inner{k_\mu}{k_\mu}}$ in
the Bloch sphere. Then $\cm$ is a mixture of von Neumann
measurements iff the set of Bloch vectors is invariant under
inversion through the centre of the sphere.

Note that we do not generally have equality in proposition
\ref{subadd}. For example if $\cm = \{\proj{k_\mu} \}$ is any POVM
in $d=2$ define the POVM $\overline{\cm}$ to be the measurement
based on subnormalised states $\ket{n_\mu}$ where $\ket{k_\mu}$
and $\ket{n_\mu}$ are orthogonal and have equal lengths. Then
generally $E_{\min }(\cm)=E_{\min }(\overline{\cm}) \neq 0$ yet
$\frac{1}{2}\cm \cup \frac{1}{2}\overline{\cm}$ has zero cost,
being a mixture of pairs of orthogonal directions i.e. a mixture
of von Neumann measurements in $d=2$.

The construction $\cm = \bigcup_i p_i \cm_i$ in proposition
\ref{subadd} corresponds to first making a classical probabilistic
choice $i$ and then applying the POVM corresponding to the seen
choice. We now consider reversing the order of these operations.
Let $\cm =\{ \proj{k_\mu} \}_{\mu=1}^m$ be any POVM and let
$\cp^\mu=\{ p^\mu_1, \ldots ,p^\mu_{K(\mu)}\}$ be a classical
probability distribution with $K(\mu)$ outcomes for each $\mu=1,
\ldots ,m$. Consider the POVM $\cm (\cp^\mu )$ defined as follows:
first apply $\cm$. If the outcome $\mu$ is seen then sample
$\cp^\mu$ to get a value $l$. Output the pair $(\mu,l)$. In
mathematical terms we take each element $\proj{k_\mu}$ of $\cm$
and replace it by $K(\mu)$ elements $p^\mu_1 \proj{k_\mu}, \ldots
, p^\mu_{K(\mu)} \proj{k_\mu}$ which are identical except for
their normalisations. Geometrically for POVMs on qubits, we
replace each Bloch vector $v_\mu$ by $K(\mu)$ vectors parallel to
$v_\mu$ whose squared lengths sum to the squared length of
$v_\mu$.

\begin{proposition}\label{clafter} For any POVM $\cm$ and
collection $\cp^\mu$ of probability distributions as above we have
\[ E_{\min }(\cm(\cp^\mu))\leq E_{\min } (\cm). \]
Thus if $\cv=\bigcup_i q_i \cv_i$ is any mixture of von Neumann
measurements then $E_{\min }(\cv (\cp^\mu))=0$. \end{proposition}

\noindent {\bf Proof}\,\, Let $B=\{ \ket{w_\mu} \}$ (with basic
ancilla state $\ket{a_0}$) be any Naimark extension of $\cm$
realising its minimal cost. For each $\mu$ let
$\ket{\alpha_1},\ldots ,\ket{\alpha_{K(\mu)}}$ be an orthonormal
set of states in a further auxiliary space with
$\inner{0}{\alpha_i}=\sqrt{p_i^\mu}$ (where $\ket{0}$ is any
choice of fixed state in the auxiliary space). To construct such a
set of states we write the $\sqrt{p_i^\mu}$'s as the first column
of a $K(\mu) \times K(\mu)$ matrix and fill out the rest of the
matrix with orthonormal columns. Then the $\ket{\alpha_i}$'s are
given (via components in a basis whose first member is $\ket{0}$)
by the rows of the matrix. With this construction, $B'=\{
\ket{w_\mu}\ket{\alpha_i^\mu} \}_{i,\mu}$ is clearly an
orthonormal set and it provides a Naimark extension of $\cm
(\cp^\mu)$ if we use basic ancilla state $\ket{a_0}\ket{0}$. Since
the ancilla has been extended by product states the entanglement
of $\ket{w_\mu}\ket{\alpha_i^\mu}$ is the same as that of
$\ket{w_\mu}$ so the entanglement cost of $\cm$ with $B$ equals
the entanglement cost of $\cm(\cp^\mu)$ with $B'$ giving the
result. $\qed$

Hence if a POVM $\cm$ can be viewed as a mixture of von Neumann
measurements followed by a classical probabilistic choice (whose
distribution may depend on the chosen von Neumann measurement's
outcome) then $\cm$ has zero cost.

\noindent {\bf Example}\,\, The above characterisation does not
generally exhaust all POVMs with zero cost. Consider the 5 outcome
POVM $\cn$ in $d=3$ defined by \[ \begin{array}{c}
\ket{k_1}=\ket{0}\hspace{5mm} \ket{k_2}=\frac{1}{\sqrt{2}}\ket{1}
\hspace{5mm}\ket{k_3}=\frac{1}{\sqrt{2}}\ket{2} \nonumber \\
\ket{k_4}=\frac{1}{2}(\ket{1}+\ket{2}) \hspace{5mm}
\ket{k_5}=\frac{1}{2}(\ket{1}-\ket{2}). \end{array}
\] One may write down a Naimark extension of product states showing
that $E_{\min} (\cn)=0$. However $\cn$ is not a mixture of Von
Neumann measurements (as the number of outcomes is not a multiple
of $d$) and it is not of the form $\cm (\cp^\mu)$ for any $\cm$ or
$\cp^\mu$ (as it has no parallel outcomes $\ket{k_\mu}$). But if
we replace the element $\ket{k_1}$ by two elements
$\ket{k_{1,1}}=\ket{k_{1,2}}=\frac{1}{\sqrt{2}}\ket{0}$ then the
resulting 6 element POVM $\cn'$ is a mixture of von Neumann
measurements defined by $\{ \ket{k_{1,1}}, \ket{k_2},\ket{k_3} \}$
and $\{ \ket{k_{1,2}},\ket{k_4},\ket{k_5} \}$. Thus $E_{\min
}(\cn')=0$ and $\cn$ arises by an identification of parallel
elements. Later (after proposition \ref{nono}) we will see that in
the special case of $d=2$, all POVMs having zero cost are of the
form $\cv(\cp^\mu)$ of proposition \ref{clafter} (i.e. where $\cv$
is a mixture of von Neumann measurements).

\begin{proposition} \label{morezero} Suppose that a POVM $\cm'=\{
\proj{k_\mu} \}_{\mu=1}^m$ has zero entanglement cost and
$\ket{k_1}$ and $\ket{k_2}$ are parallel. Let
$\proj{k_{12}}=\proj{k_1}+\proj{k_2}$. Then the rank 1 POVM $\cm
=\{ \proj{k_{12}}, \proj{k_3}, \ldots , \proj{k_m} \}$ (with $m-1$
elements) has zero cost too. \end{proposition}

\noindent {\bf Proof}\,\, Let $\{ \ket{w_\mu} \}$ with basic
ancilla state $\ket{a_0}$ be a Naimark extension of $\cm'$ with
zero cost. Thus each $\ket{w_\mu}$ for $\mu =1,\ldots ,K$ is a
product state of the form \[
\ket{w_\mu}=\ket{\hat{k}_\mu}\ket{\alpha_\mu} \] where
$\ket{\hat{k}_\mu}$ is $\ket{k_\mu}$ normalised and
$|\inner{a_0}{\alpha_\mu}|^2= \inner{k_\mu}{k_\mu}$. Since
$\ket{\hat{k}_1}=\ket{\hat{k}_2}$ we deduce that $\ket{\alpha_1}$
and $\ket{\alpha_2}$ are orthonormal. Let $\ket{\alpha_{12}}$ be
the normalised projection of $\ket{a_0}$ into the plane of
$\ket{\alpha_1}$ and $\ket{\alpha_2}$. Then
$|\inner{a_0}{\alpha_{12}}|^2
=|\inner{a_0}{\alpha_1}|^2+|\inner{a_0}{\alpha_2}|^2$.  Write
$\ket{w_{12}}=\ket{\hat{k}_1}\ket{\alpha_{12}}$. Then the
orthonormal set $\{ \ket{w_{12}}, \ket{w_3}, \ldots \}$ is a
Naimark extension for $\cm$ with zero cost. $\qed$

Above we have characterised a class of POVMs in any dimension $d$
with zero cost. Next we characterise a class that is guaranteed to
have non-zero cost and fully characterise zero cost POVMs in
$d=2$.

\begin{proposition}\label{nono}
Let $\cm=\{ \proj{k_\mu} \}_{\mu=1}^m$ be any POVM on $\ch_d$
having zero entanglement cost. Then, for any $\mu$ we have the
operator inequality
\begin{equation}
\proj{k_\mu} +\sum_{\nu: \inner{k_\mu}{k_\nu}=0}\proj{k_\nu} \ge
\inner{k_\mu}{k_\mu}{\bf 1}. \label{eq:zerocostcond}
\end{equation}
\label{prop:zerocostcond}
\end{proposition}

\noindent {\bf Proof}\,\, We may assume $\mu=1$ without loss of
generality. Let $A$ be the set of integers $\nu$ satisfying $2\le
\nu \le m$ and $\inner{k_1}{k_\nu}\neq 0$, and let $\bar{A}\equiv
\{1,\ldots, m\}-A$. Note that the LHS of
Eq.~(\ref{eq:zerocostcond}) is equal to $\sum_{\nu\in
\bar{A}}\proj{k_\nu}$. Since $\cm$ has zero entanglement cost, it
has a Naimark extension $\{\ket{w_\nu}\}_{\nu=1}^m$ in
$\ch_d\otimes \ch_e$ of the form
$\ket{w_\nu}=\ket{\hat{k}_\nu}\ket{\alpha_\nu}$, where
$|\inner{\alpha_\nu}{a_0}|^2=\inner{k_\nu}{k_\nu}$. For $\nu\in
A$, $\inner{w_1}{w_\nu}=0$ implies
$\inner{\alpha_1}{\alpha_\nu}=0$. The operator $P\equiv
\sum_{\nu\in A}\proj{w_\nu}$ is thus a projector onto a subspace
of $\ch_d\otimes (\ch_e-\ket{\alpha_1})$, and $P\le {\bf 1}\otimes
({\bf 1}-\proj{\alpha_1})$. Then we have $\sum_{\nu\in
A}\proj{k_\nu}=\bra{a_0}P\ket{a_0} \le {\bf
1}(1-|\inner{\alpha_1}{a_0}|^2)={\bf 1}(1-\inner{k_1}{k_1})$.
Since $\sum_{\nu\in \bar{A}}\proj{k_\nu}={\bf 1}-\sum_{\nu\in
A}\proj{k_\nu}$, we obtain Eq.~(\ref{eq:zerocostcond}). $\qed$

This proposition states that for each element $\ket{k_\mu}$ in a
zero-cost POVM, we can find elements orthogonal to $\ket{k_\mu}$
of a sufficient number to satisfy Eq.~(\ref{eq:zerocostcond}).
Hence we get the following corollaries.
\begin{corollary} \label{nono1} If a POVM $\cm$ has an element
$\ket{k_\mu}$ such that no other element $\ket{k_\nu}$ is
orthogonal to $\ket{k_\mu}$ then $\cm$ has non-zero entanglement
cost.
\end{corollary}

\begin{corollary}\label{nono2} Let $\cm$ be  POVM in dimension $d=2$.
Then $\cm$ has zero cost iff $\cm$ is a mixture of von Neumann
measurements followed by classical probabilistic choices i.e.
$\cm$ has the form $\cv (\cp^\mu)$ of proposition
\ref{clafter}.\end{corollary}

\noindent {\bf Proof}\,\, Suppose that $\cm$ with $d=2$ has zero
cost. By identifying all parallel sets of elements of $\cm$ we can
construct a POVM $\cm'$ satisfying
$|\inner{\hat{k}_\mu}{\hat{k}_{\mu'}}| \neq 1$ for $\mu\neq \mu'$.
By proposition \ref{morezero} $\cm'$ has zero cost too. Since
$d=2$, Proposition \ref{prop:zerocostcond} requires that any
element $\ket{k_\mu}$ in $\cm'$ should be paired to another
element $\ket{k_{\mu'}}$ satisfying $\inner{k_\mu}{k_{\mu'}}=0$
and $\inner{k_\mu}{k_\mu}=\inner{k_{\mu'}}{k_{\mu'}}$. Hence
$\cm'$ is a mixture of von Neumann measurements $\cv$ and $\cm$
has the form $\cv(\cp^\mu)$. The converse implication is given by
proposition \ref{clafter}. $\qed$

 \subsection{The entanglement cost of the trine measurement}
\label{trinemin}

The minimisation needed to determine $E_{\min } (\cm )$ cannot
generally be performed analytically except for some special POVMs.
Amongst the latter is the trine measurement (defined below), which
is perhaps the simplest generalised measurement that is not a von
Neumann measurement.

 The trine measurement is defined as a POVM
 $\{\ket{\beta_\mu}\bra{\beta_\mu}\}_{\mu=1,2,3}$  on ${\cal H}_2$,
  where
 \begin{eqnarray}
 \inner{\beta_\mu}{\beta_\mu}&=&2/3 \nonumber \\
 |\inner{\beta_\mu}{\beta_{\mu'}}|&=&1/3 \;\; {\rm for}\;\; \mu\neq
 \mu' \label{1}
 \end{eqnarray}
 The three directions $\ket{\beta_\mu}$ are equally spaced around
 a great circle on the Bloch sphere.
 Recall that this measurement is optimal for the trine source whose
 states are respectively orthogonal to the $\ket{\beta_\mu}$'s.

 \begin{theorem}\label{trinethm} The entanglement cost of the
 trine measurement is
 $\frac{2}{3}H(\frac{1}{2}(1-\frac{1}{\sqrt{3}}))= 0.496...$ ebits
 (where $H(x)= -x\log_2 x -(1-x)\log_2 (1-x)$ is the binary
 entropy function). \end{theorem}

 \noindent The proof of this theorem is given in the appendix.

\subsection{Upper bounds on the cost of any measurement $\cm$}
\label{upperbds} While determining the minimum entanglement cost
$E_{\min}({\cal M})$ for a general POVM ${\cal M}$ is difficult,
we can construct a Naimark extension having a fairly small
entanglement cost as follows.

\begin{proposition}\label{num5}
Let $\cm=\{ \proj{k_\mu} \}_{\mu=1}^m$ be any POVM on $\ch_d$.
Then, for any normalized state $\ket{\zeta}$ in $\ch_d$, there
exists a Naimark extension $\{ \ket{w_\nu} \}_{\nu=1}^{de}$ of
$\cm$ having the following form, with $e$ at most $m+1$:
\begin{equation}
 \ket{w_\nu}  =  \ket{k_\nu}\ket{a_0}+
\ket{\zeta}\ket{\xi_\nu} \;\; \mbox{for $\nu=1, \ldots ,m$}
\label{eqforpro5}
\end{equation}
where $\{ \ket{\xi_\nu} \}_{\nu=1}^{m}$ lie in the span of
$\{\ket{a_i}: i\geq 1 \}$ (denoted $\ch_e-\ket{a_0}$) and
$\{\ket{w_\nu} \}_{\nu=m+1}^{de}$ lie in
 $\ch_d \otimes (\ch_{e}-\ket{a_0})$.
This extension has entanglement cost
\begin{equation}
E= \sum_{\mu=1}^m \frac{r_\mu^2}{d} H\left(
\frac{1}{2}-\frac{1}{2}\sqrt{\alpha_\mu^2+(1-\alpha_\mu^2)
|\inner{\zeta}{\hat{k}_\mu}|^2} \right), \label{eqcost_zeta}
\end{equation}
where $r_\mu^2=\inner{k_\mu}{k_\mu}$, $\ket{\hat{k}_\mu}
=r_\mu^{-1}\ket{k_\mu}$, and $\alpha_\mu=1-2r_\mu^2$.
\end{proposition}

\noindent {\bf Proof}\,\, Existence of a Naimark extension
immediately assures us that there exists a set of $m$ vectors
$\{\ket{\tilde\xi_{\nu}}\}_{\nu=1}^{m}$ satisfying
$\inner{k_{\nu'}}{k_{\nu}}+\inner{\tilde\xi_{\nu'}}{\tilde\xi_{\nu}}
=\delta_{\nu\nu'}$. (Choose $\ket{\tilde\xi_{\nu}}\equiv
\ket{v^1_\nu}\ket{a_1}+\ldots +\ket{v_\nu^{(e-1)}}\ket{a_{(e-1)}}$
in Proposition \ref{num1}, for example.) If we take $e=m+1$, we
can then find $m$ vectors $\{\ket{\xi_{\nu}}\}_{\nu=1}^{m}$ lying
in $\ch_e-\ket{a_0}$, satisfying
$\inner{k_{\nu'}}{k_{\nu}}+\inner{\xi_{\nu'}}{\xi_{\nu}}
=\delta_{\nu\nu'}$. Using these vectors, define $\{\ket{w_\nu}
\}_{\nu=1}^{m}$ through Eq.~(\ref{eqforpro5}). The set
$\{\ket{w_\nu} \}_{\nu=1}^{m}$ is then orthonormal, and can hence
be extended to an orthonormal basis $\{\ket{w_\nu}
\}_{\nu=1}^{de}$ of $\ch_d\otimes\ch_e$ with $\{\ket{w_\nu}
\}_{\nu=m+1}^{de}$ being orthogonal to the  subspace $\ch_d
\otimes \ket{a_0}$. Proposition \ref{num1} then guarantees that
this basis is a Naimark extension of $\cm$.

The state $\ket{\omega_\nu}$ ($\nu \le m$) can be written in terms
of normalized vectors as
\[
r_\mu\ket{\hat{k}_\mu}\ket{0}+\sqrt{1-r_\mu^2}\ket{\zeta}\ket{1_\mu}
\] where $\inner{0}{1_\mu}=0$. Then the derivation of
Eq.~(\ref{eqcost_zeta}) is straightforward\cite{levitin}.  $\qed$

This construction shows that for any POVM on any size of Hilbert
space, the minimum entanglement cost is less than 1 ebit (as
$H(x)\leq 1$ and $\sum_\mu r_\mu^2/d=1$) but we can derive more
detailed upper bounds for $E_{\min}({\cal M})$.
 For any POVM, we can assume
 $\inner{k_1}{k_1}\ge \inner{k_\mu}{k_\mu} (\mu=2,\ldots,m)$ without loss
of generality. Then, choose $\ket{\zeta}=\ket{\hat{k}_1}$ in the
above construction. Since $\ket{\omega_1}$ has no entanglement and
the other $\{\ket{\omega_\nu}\}$ have less than one ebit, $E\le
\sum_{\mu=2}^m r_\mu^2/d=1-r_1^2/d$ for this construction. Noting
that $r_1^2 m\ge \sum_\mu r_\mu^2=d$, we obtain an upper bound as
a function of the number of elements $m$:
\begin{corollary}\label{thm2}
Let $\cm=\{ \proj{k_\mu} \}_{\mu=1}^m$ be any POVM on $\ch_d$.
Then,
\[
E_{\min}({\cal M})\le 1-\frac{1}{m}.
\]
\end{corollary}

We also have an upper bound as a function of $d$, by considering
the expectation value $\overline{E}$ of $E$ when $\ket{\zeta}$ is
randomly chosen  among all pure states in $H_d$. The expectation
value of $|\inner{\zeta}{\hat{k}_\mu}|^2$ is given by
$\overline{|\inner{\zeta}{\hat{k}_\mu}|^2}=1/d$. We will use the
following lemma.
\begin{lemma}
$H[(1-\sqrt{x})/2]$ is a concave function of $x$ for $0\le x \le
1$. \label{lemconcavity}
\end{lemma}
\noindent {\bf Proof}\,\, For $0 \le x <1$, the first derivative
of this function  can be expanded as \[ -(\ln
4)^{-1}\sum_{m=0}^{\infty} x^m/(2m+1) \] which is obviously
monotone decreasing. $\qed$

Using this lemma and Eq.~(\ref{eqcost_zeta}), we have
\[
\overline{E}\le \sum_{\mu=1}^m \frac{r_\mu^2}{d} H\left(
\frac{1}{2}-\frac{1}{2}\sqrt{\alpha_\mu^2+(1-\alpha_\mu^2)
\overline{|\inner{\zeta}{\hat{k}_\mu}|^2}} \right) \le
\sum_{\mu=1}^m \frac{r_\mu^2}{d}
H\left[\frac{1}{2}\left(1-\frac{1}{\sqrt{d}}\right)\right].
\]
Since $E_{\min}({\cal M})$ is not larger than this averaged value,
we obtain the following upper bound.

\begin{theorem}\label{thm3}
Let $\cm=\{ \proj{k_\mu} \}_{\mu=1}^m$ be any POVM on $\ch_d$.
Then,
\[
E_{\min}({\cal M})\le
H\left[\frac{1}{2}\left(1-\frac{1}{\sqrt{d}}\right)\right].
\]
\end{theorem}

Some values of this upper bound are given in the following table.
\[ \begin{array}{ccc} d & &
H\left[\frac{1}{2}\left(1-\frac{1}{\sqrt{d}}\right)\right]
\nonumber \\ & & \nonumber \\
 2 &  & 0.600876 \nonumber \\ 3 & & 0.744008=0.469417\log_2 3 \nonumber \\
 4 & & 0.811278=0.405639\log_2 4 \nonumber \end{array} \]

\section{The most costly generalised measurement} \label{mostcostly}

To any (discrete) POVM $\cm$ we have assigned its minimal
entanglement cost $E_{\min } (\cm )$ and it is interesting to ask:
which POVM $\cm$ (for any $d$) has the {\em largest} such cost?
Furthermore we may prefer to scale the cost by $\log_2 d$, the
number of qubits in the source space and ask for $\cm$ that
maximises the ``normalised entanglement cost'' $F(\cm)=E_{\min
}(\cm)/\log_2 d$ i.e. the POVM which requires the maximal
entanglement investment per qubit of source support.

According to theorem \ref{thm3}, for source dimension $d$ we have
\[ F(\cm)\leq \frac{1}{\log_2 d}
H\left[\frac{1}{2}(1-\frac{1}{\sqrt{d}})\right]. \] This upper
bound is a decreasing function of $d$ with value $0.46...$ for
$d=3$. Recalling that the trine measurement has minimum cost
$0.49...$ in $d=2$ we see that $F(\cm )$ is maximal for $d=2$ and
$\cm$ must have entanglement cost at least as large as the trine
measurement. Although we have been unable to exhaustively search
the POVM space of $d=2$ (either numerically or analytically), we
do have a numerical scheme for evaluating the entanglement cost of
any POVM in $d=2$ under the restriction that the ancilla has
dimension 2. Using this scheme we have found that no other POVM
with $d=2$ and $e=2$ has a cost as large as the trine cost.
Details of the numerical scheme and further numerical studies will
be given in a forthcoming publication\cite{presnell}.

Hence we have shown:
\begin{theorem}\label{trinemax} The trine measurement is the
unique POVM with maximal normalised entanglement cost $E_{\min }
(\cm )/\log_2 d$ amongst all POVMs with \\ (a) $d\geq 3$, or \\
(b) $d=2$ and ancilla size $e=2$. \end{theorem} This provides
strong evidence for\\ {\bf Conjecture}\, The trine measurement is
the unique POVM with maximal normalised entanglement cost, amongst
all (discrete) POVMs.

To decide the conjecture it remains only to treat the case of
$d=2$ with $e\geq 3$ (including POVMs with an unboundedly large
number of outcomes).

\section{Asymptotic entanglement cost}\label{asymptcost}

So far we have been considering the entanglement cost of a single
instance of a POVM $\cm$. In quantum information theory it is
often advantageous and interesting to consider an asymptotic
scenario in which information is collected from more and more
instances of the target system and we assess the minimal cost per
system that results from allowing global operations on all
instances jointly. In this spirit, for a POVM $\cm$ we may
consider the entanglement cost of $n\cm$, the POVM defined by
applying $\cm$ $n$ times. We introduce the {\em asymptotic
entanglement cost of $\cm$} as being $\lim_{n\rightarrow \infty}
E_{\min} (n\cm)/n$. If we implement $n\cm$ simply as $n$
independent instances of $\cm$ then the resulting entanglement
cost per measurement will be $nE(\cm)/n=E(\cm)$. However if we
consider global Naimark extensions of the single POVM defined by
$n\cm$ on $\ch_{d}^{\otimes n}$ then theorem \ref{thm3}
immediately gives:

\begin{theorem} \label{asympt} For any POVM the asymptotic
entanglement cost is zero. \end{theorem}

\noindent Further discussion of the significance of this result is
given below in section \ref{entorspace}.

\section{Concluding remarks}\label{etc}
\subsection{Other formulations?}\label{others}

Our notion of entanglement cost -- based on the entanglements of
the Naimark extension states -- is ``retrospective'' in the sense
that it quantifies the entanglement {\em produced} by the
measurement. In contrast it would be interesting to develop a
quantitative notion of ``active'' cost i.e. the entanglement {\em
input} required to {\em perform} the POVM, in some suitable sense.
A similar situation occurs in the question of quantifying the
amount of entanglement in a bipartite mixed state. We have $E_D$,
the entanglement of distillation i.e. the amount of entanglement
that can be produced, analogous to our notion of POVM cost, and
$E_F$, the entanglement of formation i.e. the amount of
entanglement that needs to be invested to prepare the state. Just
as $E_D\leq E_F$ for all states, we would expect that our $E_{\min
}(\cm )$ would be  a lower bound on any reasonable notion of input
entanglement cost required to implement the POVM.

There are various possible approaches to formulating a notion of
an input entanglement cost. For example, we could demand that the
only allowed measurement is a
   standard product von Neumann measurement of the form $\{
   \ket{e_j}\ket{l} \}$ on $\ch_d\otimes \ch_e$ (for any desired
   ancilla space with orthonormal basis $\{ \ket{l} \}$).
   Then to implement a POVM $\cm$ we
   first apply a unitary transformation $U$ on $\ch_d\otimes\ch_e$ so
   that the measurement of $\cm$ on $\ket{\psi_i}\ket{a_0}$ is
   equivalent to the product measurement on $\ket{\tilde{\psi}_i}=
   U\ket{\psi_i}\ket{a_0}$. Finally we compute the entanglement $E_i$
   of $\ket{\tilde{\psi}_i}$  and ask for
   the minimal average $E=\sum p_i E_i$, minimised over all choices
   of Naimark extension for $\cm$.
   The average here is constructed using
   the prior probabilities of the signal states.
   Although this approach is appealing -- in that we explicitly
   consider the amount of entanglement $E_i$ that needs to be ``put
   into'' the source states -- it is unfortunately trivial as the
   answer is always $E_i^{\rm (min)}=0$! To see this, note that we
   can always begin the action of $U$ on $\ket{\psi_i}\ket{a_0}$ by
   swapping the signal state out entirely into the ancilla and
   leaving the signal space in a standard state $\ket{s_0}$
   (independent of $i$). Then any generalised measurement on the
   source can be simulated by subsequent actions entirely in $\ch_e$
   with the system plus ancilla always being in a product state,
   having $\ket{s_0}$ in the signal space i.e. $E_i =0$. (This kind
   of trivial entanglement--avoiding construction was also noticed in
   \cite{busch}).

   To avoid this problem of the minimal $E$ being identically zero,
   we could instead ask about the ``amount of coupling'' needed to
   implement the unitary operation $U$ on $\ch_d\otimes
   \ch_e$.  There are a number of possible meanings to
   ``amount of coupling'', and this is a subject of considerable current
interest
    \cite{operations}. For example one can talk about the maximum amount
of entanglement that can be produced by $U$ or the amount of
entanglement needed to implement $U$.  Or one can consider how
long it takes to implement $U$ using a fixed Hamiltonian.
   Interestingly, in the context of the above discussion, the swap
operation on $\ch_d\otimes
   \ch_d$ typically has a {\em high} cost for these quantities. Thus the
method in the
   previous paragraph for having $E_i=0$ (making essential use of
a swap operation) would become expensive (and would presumably no
longer be a minimal strategy). However progress in understanding
non-local aspects of bi-partite interactions has not yet reached
the stage where the possible measures of interaction are well
understood, or indeed straightforward to calculate for typical
operations. Therefore further development of this approach must
await the introduction of a (suitably tractable) definition of
interaction cost for bipartite transformations.

Although we do not have a general notion of entanglement cost of a
bipartite unitary operation, it is interesting to note that a
definition of entanglement cost for a POVM, based on such a
notion, would be expected to be {\em inequivalent} to the
definition adopted in this paper. To see this consider the product
basis of two qubits given by:
\[ \ket{\psi_0}= \ket{0}\ket{0} \hspace{5mm}
\ket{\psi_1}=\ket{1}\ket{0} \hspace{5mm} \ket{\psi_2}=
\ket{+}\ket{1} \hspace{5mm} \ket{\psi_3}=\ket{-}\ket{1} \] where
$\ket{\pm}=\frac{1}{\sqrt{2}}(\ket{0}\pm\ket{1})$. If we take the
initial ancilla state to be $\ket{a_0}=\ket{+}$ then this product
basis is actually a Naimark extension of the 4 element POVM
$\{\proj{k_\mu}\}$ with \begin{equation} \label{kpom}
\ket{k_1}=\frac{1}{\sqrt{2}} \ket{0}\hspace{5mm}
\ket{k_2}=\frac{1}{\sqrt{2}}
\ket{1}\hspace{5mm}\ket{k_3}=\frac{1}{\sqrt{2}}
\ket{+}\hspace{5mm}\ket{k_4}=\frac{1}{\sqrt{2}} \ket{-}.
\end{equation} (This POVM may be interpreted as the 50/50
probabilistic mixture of two von Neumann measurements $\{ \ket{0},
\ket{1} \}$ and $\{ \ket{+}, \ket{-} \}$: we first measure the
ancilla $\ket{a_0}$ in the basis $\{ \ket{0}, \ket{1} \}$ (cf the
second qubit of the $\ket{\psi_i}$ states) and then we choose the
von Neumann measurement for the first qubit according to whether
the outcome is  0 or 1.) Since we have a Naimark extension of
product states this POVM has zero entanglement cost by our adopted
definition. However it can be shown that no product unitary
transformation on two qubits can transform the bi-orthogonal
product basis $\{ \ket{j}\ket{l} \}_{j,l=0}^1$ to the basis $\{
\ket{\psi_i} \}_{i=0}^3$. i.e. the POVM would be expected to
require interaction between system and ancilla and thus have
non-zero entanglement cost in the new sense. It is interesting to
note that if, as a form of interaction, we allow free classical
communication between system and ancilla then the POVM in eq.
(\ref{kpom}) again has zero cost (using the protocol described
after eq. (\ref{kpom}), with the ancilla measurement result being
communicated to the system).

\subsection{Entanglement or space?}\label{entorspace}

Although we have introduced a notion of entanglement cost for
POVMs, theorem \ref{asympt} indicates that in an asymptotic sense,
entanglement between system and ancilla is not important since the
(suitably normalised) average entanglement of the measurement
outcomes is zero.  This leads to the question of what role the
ancilla actually plays i.e. we wish to characterise a resource
that is being consumed. If the number $M$ of POVM elements exceeds
the dimension $d$ of the source space, then in the asymptotic
scenario of $n$ applications of $\cm$, the Naimark extension
requires $M^n$ orthogonal directions which is exponentially larger
than the source dimension $d^n$. Here the ancilla provides the
extra space. Hence we are lead to the intriguing suggestion that
``unentangled space'' itself might be a quantifiable and useful
resource in quantum information processing. Further exploration of
this idea will be given in a future paper.

One may also consider a similar asymptotic formulation in the
context of extracting information from sources. Let $\ce$ be a
source in dimension $d$ and let $\ce^{\otimes n}$ be the source
(in dimension $d^n$) whose signal states are blocks of $n$
independent emissions from $\ce$ (with prior probabilities for
$\ce^{\otimes n}$ being products of those from $\ce$). The
accessible information $\acc (\ce^{\otimes n})$ of $\ce^{\otimes
n}$ is determined by optimising the mutual information over all
POVMs in dimension $d^n$ and it is known that $\acc (\ce^{\otimes
n}) = n\acc (\ce )$. But to focus on the role of the ancilla we
introduce firstly, $\acc^\perp (\ce^{\otimes n})$, the maximal
mutual information obtainable by applying only von Neumann
measurements in dimension $d^n$ (i.e. no ancilla needed) and
secondly, $\acc^{0} (\ce^{\otimes n})$, the maximal mutual
information obtainable by applying only POVMs in dimension $d^n$
that have zero entanglement cost (i.e. an ancilla is needed but it
remains unentangled with the source). Then we can ask: in the
limit of large $n$, does either $\acc^\perp (\ce^{\otimes n})/n$
or $\acc^{0} (\ce^{\otimes n})/n$ approach $\acc (\ce )$? In
\cite{massarpopwin} a related question has been considered, where
the object function is not the mutual information but a measure of
success probability of identifying some desired property of the
signal states. In this context it was shown that we may choose
optimal POVMs that resemble von Neumann measurements more and
more, in the asymptotic limit.

\section{Appendix: proof of theorem \ref{trinethm}}

Any Naimark extension $\{w_\nu\}_{\nu=1}^{de}$ of the trine POVM
 $\{\ket{\beta_\mu}\bra{\beta_\mu}\}_{\mu=1,2,3}$ in ${\cal H}_2
 \otimes {\cal H}_e$ has the following form
 $$
 \ket{w_\mu}=\ket{\beta_\mu}\ket{a_0}+
 \sqrt\frac{1}{3}\ket{\chi_\mu} \;\; {\rm for} \;\; \mu=1,2,3,
 $$
 where $\{\ket{\chi_\mu}\}_{\mu=1,2,3}$ and
 $\{\ket{w_\nu}\}_{\nu=4}^{de}$ lie in $\ch_2\otimes
 (\ch_e-\ket{a_0})$. The orthonormality of $\{ \ket{w_\nu}
 \}_{\nu=1}^{3}$ leads to $|\inner{\chi_\mu}{\chi_{\mu'}}|=1$ for
 any $1\le \mu, \mu'\le 3$, implying that
 $\{\ket{\chi_\mu}\}_{\mu=1,2,3}$ are the same except for their
 phase factors, and the reduced state of $
 \ket{\chi_\mu}\bra{\chi_\mu}$ on $\ch_2$ is independent of $\mu$.
 Let us take an orthonormal basis $\{\ket{0},\ket{1}\}$ of $\ch_2$
 such that this common density operator
 $\tr_e( \proj{\chi_\mu} )$, having rank 2, is given by
 $p\ket{0}\bra{0}+[(1-p)/2]\bm{1}$.

The average entanglement $E$ for this extension is
 $$
 E=\sum_\mu \frac{1}{3} S(\rho_\mu),
 $$
 where
 $$
 \rho_\mu\equiv \frac{2}{3}\ket{\hat\beta_\mu}\bra{\hat\beta_\mu} +
 \frac{p}{3}\ket{0}\bra{0} +\frac{1-p}{6}\bm{1}
 $$
 and $\ket{\hat\beta_\mu}$ is $\ket{\beta_\mu}$ normalized.
 Let us first consider the dependence of $E$ on $p$, with
$\{\ket{\beta_\mu}\}$ fixed. If we define
 density operators $\tilde{\rho}_\mu^{(p)}$ by
 $$
 \tilde{\rho}_\mu^{(p)}\equiv \frac{2}{3}\ket{0}\bra{0} +
 \frac{p}{3}\ket{\hat\beta_\mu}\bra{\hat\beta_\mu} +\frac{1-p}{6}\bm{1},
 $$
 it is obvious that $S(\rho_\mu)=S(\tilde{\rho}^{(p)}_\mu)$.
Noting that
$$
\rho_1^{(p)}=\frac{1+2p}{3}\rho_1^{(1)}+\frac{1-p}{3}\rho_2^{(1)}
+\frac{1-p}{3}\rho_3^{(1)}
$$
and similar expressions for $\rho_2^{(p)}$ and $\rho_3^{(p)}$, we
obtain $\sum_\mu S(\tilde{\rho}^{(p)}_\mu)\ge \sum_\mu
S(\tilde{\rho}^{(1)}_\mu)$ from the concavity of $S$. Hence, in
finding the minimum of $E$, we only
 need to consider the case $p=1$. In the following we assume that
 $p=1$.

 Let us introduce Pauli operators on $\ch_2$ defined by
 $\sigma_x\equiv\ket{1}\bra{0}+\ket{0}\bra{1}$, $\sigma_y\equiv
 i\ket{1}\bra{0}-i\ket{0}\bra{1}$, and $\sigma_z\equiv
 \ket{0}\bra{0}-\ket{1}\bra{1}$. Define Bloch vectors for
 $\ket{\hat{\beta}_\mu}$ by $\bm{p}^{(\mu)}\equiv
 \bra{\hat{\beta}_\mu}\bm{\sigma}\ket{\hat{\beta}_\mu}$, where
 $\bm{\sigma}=(\sigma_x,\sigma_y,\sigma_z)$. The eigenvalues of
 $\rho_\mu$ are then given by
 $$
 \frac{1}{2}\pm \frac{\sqrt{4p_z^{(\mu )}+5}}{6}
 $$
 where
 $$
 p_z^{(\mu)}= 2|\inner{0}{\hat{\beta}_\mu}|^2-1.
 $$
 We can write $E$ as a function of
 $(p_z^{(1)},p_z^{(2)},p_z^{(3)})$ as
 $$
 E(p_z^{(1)},p_z^{(2)},p_z^{(3)})=\sum_\mu \frac{1}{3}
 f(p_z^{(\mu)}),
 $$
 where
 $$
 f(z)\equiv H\left(\frac{1}{2}+ \frac{\sqrt{4z+5}}{6}\right).
 $$
 As can be seen from Lemma \ref{lemconcavity} (in section \ref{upperbds}),
  $f(z)$ is concave
 and hence $E(p_z^{(1)},p_z^{(2)},p_z^{(3)})$ is concave as a
 function of the triplet $(p_z^{(1)},p_z^{(2)},p_z^{(3)})$.

 Next, let us consider the possible range of the triplet
 $(p_z^{(1)},p_z^{(2)},p_z^{(3)})$. From Eq.~(\ref{1}), the Bloch
 vectors $\bm{p}^{(i)}$ have unit length and they are equally
 spaced around a great circle on the Bloch sphere. Hence we can
 derive the following:
 \begin{eqnarray}
 \bm{p}^{(1)}+\bm{p}^{(2)}+\bm{p}^{(3)}=\bm{0} \label{2}
 \\
 |\bm{p}^{(\mu )}|^2=1
 \\
 |\bm{p}^{(1)}-\bm{p}^{(2)}|^2=2-2\cos(2\pi/3)=3
 \\
 (\bm{p}^{(1)}-\bm{p}^{(2)})\cdot \bm{p}^{(3)}=0
 \end{eqnarray}
 We will use the following Lemma.
 \begin{lemma}  If $\bm{p}\cdot\bm{q}=0$,
 $$
 \frac{p_z^2}{|\bm{p}|^2}+\frac{q_z^2}{|\bm{q}|^2}\le 1.
 $$
 \end{lemma}

 \noindent {\bf Proof}\,\, Let us decompose each of the two vectors
 into the $z$ component and the $xy$-plane component {\em viz.}
 $\bm{p}=p_z \bm{z} + \bm{p}_\perp$ and $\bm{q}=q_z \bm{z} +
 \bm{q}_\perp$. Then, $\bm{p}\cdot\bm{q}=p_zq_z+\bm{p}_\perp\cdot
 \bm{q}_\perp=0$, implying that $(p_zq_z)^2\le
 |\bm{p}_\perp|^2|\bm{q}_\perp|^2
 =(|\bm{p}|^2-p_z^2)(|\bm{q}|^2-q_z^2)$. $\qed$

 Applying this Lemma to $\bm{p}^{(1)}-\bm{p}^{(2)}$ and
 $\bm{p}^{(3)}$, we have
 \begin{equation}
 (p_z^{(1)}-p_z^{(2)})^2/3+(p_z^{(3)})^2\le 1. \label{6}
 \end{equation}
 From Eq.~(\ref{2}), we also have
 \begin{equation}
 p_z^{(1)}+p_z^{(2)}+p_z^{(3)}=0. \label{7}
 \end{equation}
 If we change the variables by
 $$
 \left(
 \begin{array}{c}
 X\\Y\\Z
 \end{array}
 \right) = \left(
 \begin{array}{ccc}
 \frac{1}{\sqrt{2}}&-\frac{1}{\sqrt{2}}&0\\
 \frac{1}{\sqrt{6}}&\frac{1}{\sqrt{6}}&-\frac{2}{\sqrt{6}}\\
 \frac{1}{\sqrt{3}}&\frac{1}{\sqrt{3}}&\frac{1}{\sqrt{3}}
 \end{array}
 \right) \left(
 \begin{array}{c}
 p_z^{(1)}\\p_z^{(2)}\\p_z^{(3)}
 \end{array}
 \right),
 $$
 then the conditions (\ref{7}) and (\ref{6}) become
 \begin{eqnarray}
 Z=0
 \\
 X^2+Y^2\le 3/2.
 \end{eqnarray}
 The region specified above is a convex set, with the extreme
 points forming a circle. Note that relabeling of $(1,2,3)$ in
 Eq.~(\ref{6}) merely reproduces the same conditions. Let us
 parameterize the extreme points by an angle $\theta$, such that
 $X=\sqrt{3/2}\cos(\theta+\pi/6)$,
 $Y=\sqrt{3/2}\sin(\theta+\pi/6)$, and $Z=0$. Then $E$ is written
 as
 $$
 E(\theta)=\frac{1}{3}\sum_{k=-1,0,1}f\left( \cos(\theta+\frac{2
 k\pi}{3}) \right)
 $$
 Since $E(\theta+2\pi/3)=E(\theta)$ and $E(-\theta)=E(\theta)$,
 $$
 \min_\theta E(\theta)=\min_{0\le\theta\le \pi/3} E(\theta).
 $$
 Now we can numerically check that $E'(\theta)\le 0$ for
 $0\le\theta\le \pi/3$. This implies
 $$
 \min_{0\le\theta\le \pi/3} E(\theta)=E(0)=
 \frac{2}{3}H\left( \frac{1}{2}(1-\frac{1}{\sqrt{3}})\right)
  =0.496..
 $$
 Since $E(p_z^{(1)},p_z^{(2)},p_z^{(3)})$ is concave,
 $$
 \min E(p_z^{(1)},p_z^{(2)},p_z^{(3)})\ge\min_\theta E(\theta)
 $$

 Since there exists a valid point $(p_z^{(1)},p_z^{(2)},p_z^{(3)})$
 corresponding to $\theta=0$ (e.g., take
 $\ket{\psi}=\ket{\hat{\beta}_1}$), we conclude that
 $$
 \min E=\frac{2}{3}H\left( \frac{1}{2}(1-\frac{1}{\sqrt{3}})\right) =0.496..
 \hspace{1cm} \qed $$

 \noindent {\Large\bf Acknowledgements}

 \noindent
RJ and AW are supported by the U.K. Engineering and Physical
Sciences Research Council. SP is supported by the U.K. Engineering
and Physical Sciences Research Council and the U.K. Government
Communications Head Quarters.


\begin{thebibliography}{10}
\bibitem{peres} Peres, A. (1993) {\em Quantum theory: concepts and
methods} (especially \S9.6)  Kluwer academic publishers.

\bibitem{naimarkthm} Naimark, M. A. (1940) {\em Izv. Akad. Nauk
SSSR,  Ser. Mat.} {\bf 4}, p277-318.

\bibitem{davies} Davies, E. B. (1978) {\em IEEE Trans. Inform.
Theory} {\bf IT24}, p596-599.

\bibitem{levitin} Levitin, L. (1995) {\em Proc. Quantum
communication and measurement}, ed. V. Belavkin, O. Hirota and R.
Hudson, p439-448, Plenum Press, New York.

\bibitem{holevo} Holevo, A. (1973) {\em Prob. Peredachi Inform.} {\bf 9}
no. 2, p31-42.

\bibitem{sasakietc} Sasaki, M., Barnett, S., Jozsa, R., Osaki, M.
and Hirota, O. (1999) {\em Phys. Rev.} {\bf A59}, p3325-3335.

\bibitem{preskill} Preskill, J. (1998) {\em Lecture Notes for
Physics 229: Quantum Information and Computation}, California
Institute of Technology.

\bibitem{busch} Busch, P. (2002) {\em Note on non-entangling measurements},
preprint available at quant-ph/0209090.

\bibitem{presnell} Presnell, S. (2003) Ph.D. thesis, in
preparation.

\bibitem{operations} Cirac, J.I.,  D\"ur, W.,  Kraus, B. and  Lewenstein,
M.(2001) {\em Phys Rev Lett} {\bf 86}, 544; Collins, D.,  Linden,
N. and  Popescu, S. (2001) {\em Phys Rev} {\bf A 64}, 032302;
 Eisert, J.,  Jacobs, K.,  Papadopoulos, P. and  Plenio, M. (2001)
 {\em Phys Rev} {\bf A 62},
052317; Leifer, M.,  Henderson, L. and  Linden, N. (2003) {\em
Phys Rev} {\bf A67}, 012306; Bennett, C.,  Harrow, A.,  Leung, D.
and  Smolin, J. (2002) quant-ph/0205057; Khaneja, N.,  Brockett,
R. and  Glaser, S. (2001) {\em Phys Rev} {\bf A63}, 032308;
Bennett, C.,  Cirac, J. I.,  Leifer, M.,  Leung, D.,
  Linden, N.,
Popescu, S. and  Vidal, G. (2002) {\em Phys Rev} {\bf A66},
012305.


\bibitem{massarpopwin} Massar, S. and Popescu, S. (2000) {\em Phys. Rev.}
{\bf A61}, p062303; Winter, A. and Massar, S. (2001) {\em Phys.
Rev.} {\bf A64}, p012311.


\end{thebibliography}
\end{document}